\documentclass[prb,aps, twocolumn,showpacs]{revtex4}
\usepackage{graphicx}
\newcommand{\be}{\begin{equation}}
\newcommand{\ee}{\end{equation}}
\newcommand{\bea}{\begin{eqnarray}}
\newcommand{\eea}{\end{eqnarray}}

\newcommand{\Euler}{\gamma_{\rm\scriptscriptstyle E}}

\begin{document}

\title{Phase Transitions in Isolated Vortex Chains}

\author{
  Matthew J.\ W.\ Dodgson
}

\affiliation{Theory of Condensed Matter Group,
Cavendish Laboratory, Cambridge, CB3 0HE, UK.}

\date{January 11, 2002} 
\begin{abstract}

In very anisotropic layered superconductors (e.g.\ Bi$_2$Sr$_2$CaCu$_2$O$_x$)
a tilted magnetic field can penetrate as two 
co-existing lattices of vortices parallel and perpendicular to the layers. 
At low out-of-plane fields the perpendicular vortices form a set of isolated 
vortex chains,
which have recently been observed in detail with scanning Hall-probe 
measurements.
We present calculations that show a very delicate stability
of this  isolated-chain state. As
the vortex density increases along the chain there is a first-order
transition to a buckled chain, and
 then the chain
will expel vortices in a continuous transition to a composite-chain state.
At low densities there is an instability towards clustering, due to a 
long-range attraction between the vortices on the chain, and at very low
densities it becomes energetically favorable to
form a tilted chain, which may explain the sudden disappearance
of vortices along the chains seen in recent experiments.

\end{abstract}


\maketitle


\section{Introduction}\label{sec:intro}

The vortex system in layered
superconductors,\cite{Blatterreview} which includes the high-$T_c$ cuprates,
displays a rich set of physical phenomena, such as a thermodynamic melting
transition,\cite{Zeldov}
a pinning-induced disordering transition,\cite{vanderBeek}
and various structural transitions at different angles.\cite{Mirkovic,Ooi2000}
The average density and orientation of the vortices are controlled by the
magnetic field, because each vortex carries one quantum of flux, 
$\Phi_0=hc/2e$. 
In this paper we
are concerned with the vortex chains that appear in a certain regime of
tilted magnetic fields. These chains, which consist of a high density of 
flux lines perpendicular to the layers, were first observed with the
Bitter decoration technique by Bolle et al.\cite{Bolle} A qualitative
explanation followed shortly\cite{Huse} in terms of the proposed
crossing-lattice state in tilted fields.\cite{Theodorakis,Bulaevskii92}
This state consists
of a lattice of flux lines perpendicular to the layers crossed by a lattice
of flux lines along the layers. The in-plane lattice is strongly distorted
due to the anisotropy,\cite{Campbell} and has a large spacing
along the layers. Huse surmised that a possible attractive interaction between
the two species of flux line would lead to a higher density of out-of-plane
flux lines along chains, with an inter-chain separation equal to the in-plane
flux-line spacing. This picture seems to be consistent with the experimental
observations.\cite{Grigorieva}
The energy of the crossing-lattice state was considered by 
Benkraouda and Ledvij,\cite{Benkraouda} who found a transition from a single
lattice of tilted flux lines to the crossing-lattice state as the tilt angle
is increased for sufficiently large anisotropy. Their work, however, neglected
interactions between the two crossing lattices.

Interest in the crossing-lattice state has been revitalized in the last 
couple of years. Koshelev\cite{Koshelev99} has shown that the regime for 
crossing lattices is larger than previously expected. This is because a
correct treatment must include the interactions between the perpendicular
flux lines, and a distortion away from ideal crossing lattices has a lower 
energy. Experimental evidence comes in the form of the unusual dependence
of the melting temperature of the vortex lattice as a function of magnetic
field angle,\cite{Schmidt,Ooi1999} 
which is consistent with the crossing-lattice
state rather than a single tilted flux-line lattice.\cite{Koshelev99}
In addition, the work of Koshelev explained quantitatively
the attraction of perpendicular flux lines that leads to the vortex-chain state
in a certain field range.
Apart from more detailed work on the melting transition,\cite{Mirkovic}
and on magnetization curves,\cite{Ooi2000}
these concepts have also inspired recent scanning Hall-probe 
measurements\cite{Grigorenko}
on the vortex-chain state. These experiments are similar to the Bitter 
decoration technique in that they only probe the field distribution emanating
from the surface of the superconductor. However, they have the advantage of
speed and resolution, and are not a ``one-off'' measurement, so that the
system can be finely tuned to observe different effects. These new experiments
have shown unexpected properties within the chain state. In particular, it is
possible to tune to a field range where all of the out-of-plane flux lines
are arranged in chains. It is this ``isolated-chain state'' that we will
study in this paper.

We will use the following geometry. The $z$-axis is perpendicular to the
layers, and the component of magnetic induction $B_z$ determines the density
of ``pancake'' vortices\cite{Pancakes} in every superconducting layer. 
The in-plane field
is along the $x$-axis, and $B_x$ gives the density of the flux lines in this
direction. These in-plane flux lines have their centers in the spacing between
layers, and so are commonly called Josephson vortices.\cite{Jvortex}
A $z$-directed flux line is made from a stack of pancake vortices, and contains
circulating currents in the layers up to a distance of the penetration 
depth, $\lambda_{\rm ab}\equiv\lambda$,
from the vortex center. A Josephson vortex in the $x$-direction has an 
ellipsoidal current pattern, with the flux and currents decaying over the
much larger distance $\lambda_{\rm c}=\gamma\lambda$ 
in the $y$-direction. Here, $\gamma$ is
the anisotropy ratio, which is large for weakly coupled layers. The phase 
singularity of the Josephson vortex is confined between two neighboring 
layers, with separation $d$, where the phase difference across the layers 
changes by $2\pi$ over a distance of the Josephson length $\gamma d$.

\begin{figure}
\vspace{-0.3cm}
\centerline{\resizebox{8cm}{!}{\includegraphics{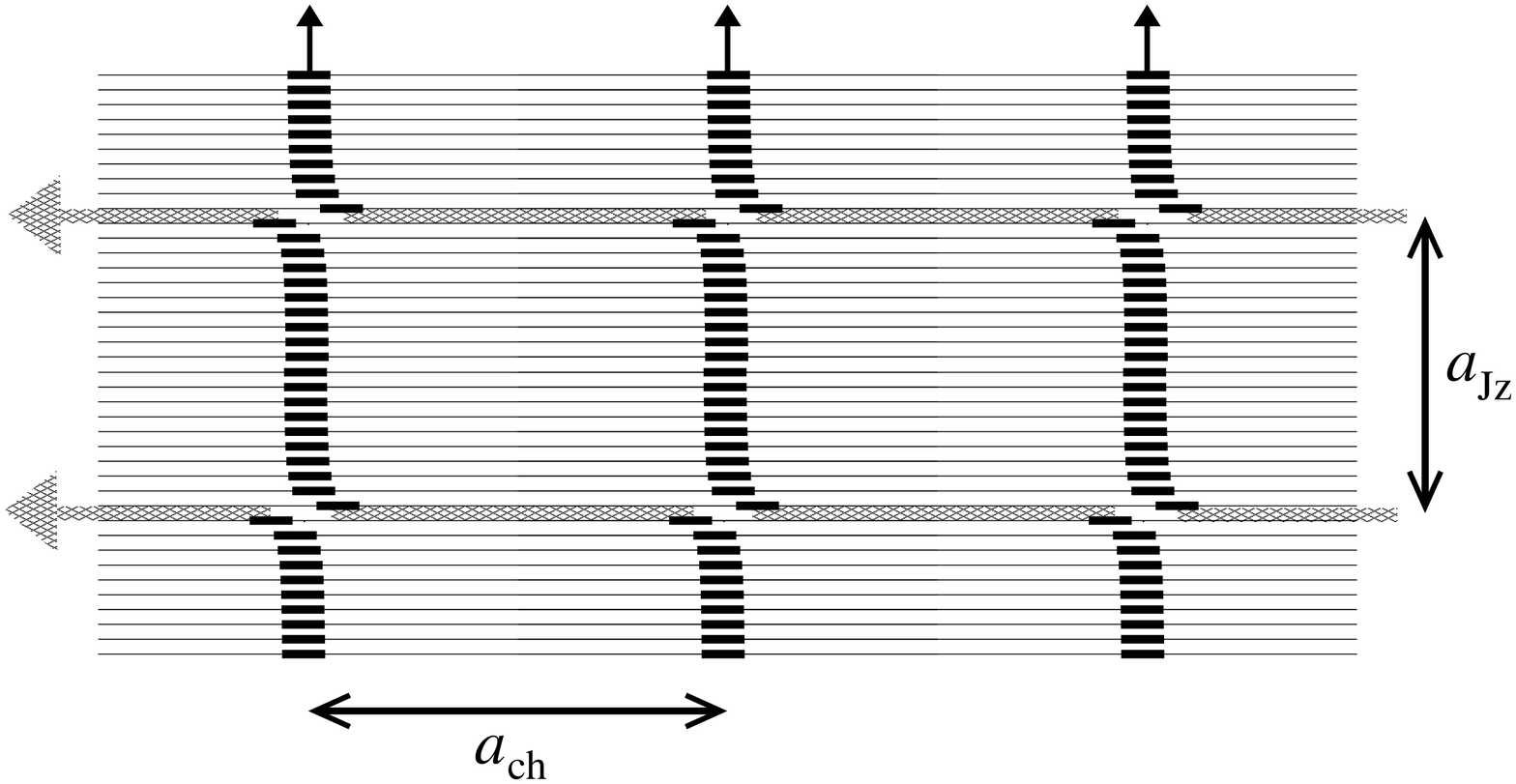}}}
\caption{ Optimal displacements of the vortex chain of pancake stacks due
to the crossing of Josephson vortices, using the parameters appropriate for
BiSCCO quoted in the text. The figure takes the results from
calculations for $a_{\rm ch}=10\lambda$, and $a_{{\rm J}z}=20 d$ 
which for the isolated-chain state corresponds to 
$B_x=53$~G and $B_z=0.8$~G. The displacements are
magnified by two for clarity, and the $z$-scale is not the same as the 
$x$-scale.
  }
\label{fig:chain}
\end{figure}
The most simple structure of the crossing-lattice state is when an ideal  
triangular lattice (with spacing $a_{\rm P}=[(2/\sqrt{3})\Phi_0/B_z]^{1/2}$)
of pancake-vortex stacks crosses a stretched triangular lattice of Josephson
vortices, with separation in the $z$-direction of $a_{{\rm J}z}=
[(2/\sqrt{3})\Phi_0/B_x]^{1/2}/\sqrt{\gamma}$.
Koshelev has shown how the interactions between pancake stacks and
Josephson vortices lead to an effective attraction.\cite{Koshelev99}
At the low pancake
densities we are interested in here ($B_z<\Phi_0/\lambda^2$) this means
that there will be large distortions from the ideal triangular pancake-vortex
lattice, with a higher density of pancake stacks located over the centers
of the Josephson vortices. This leads to the high-density chains of pancake 
stacks that are observed by Bitter decoration,\cite{Bolle,Grigorieva}
 with an inter-chain spacing of $a_{{\rm J}y}=
\sqrt{\gamma}[(\sqrt{3}/2)\Phi_0/B_x]^{1/2}$.
Eventually, for small $B_z\ll\Phi_0/\lambda^2$, 
all of the pancakes lie over the Josephson 
vortices. This isolated-chain state has a pancake-stack separation along
the chains of $a_{\rm ch}=(\Phi_0/B_z)/a_{{\rm J}y}$, see Fig.~\ref{fig:chain}.

The crossing lattices must compete energetically with the more conventional
tilted lattice of vortices. This is quite similar to the vortex lattice in a
continuous superconductor, only there is a kinked structure along each vortex
with a periodicity $L=d(B_x/B_z)$. The anisotropy of the currents mean that 
there is a stretched aspect ratio of this lattice. In addition, this distortion
may be different from that expected from simple rescaling\cite{BlatterGL}
due to the attraction of tilted vortices along the direction of tilt. This
may lead to a ``tilted chain state'',\cite{TiltedChains} which is distinct
from the crossing chains mentioned above.
Koshelev showed that for intermediate pancake densities $B_z>\Phi_0/\lambda^2$
there should be a transition at a small value of $B_x=B_x^*\approx 0.01B_z$ 
between the tilted lattice ($B_x<B_x^*$) and the crossing lattices
($B_x^*<B_x$).

\begin{figure}
\centerline{\resizebox{8cm}{!}{\includegraphics{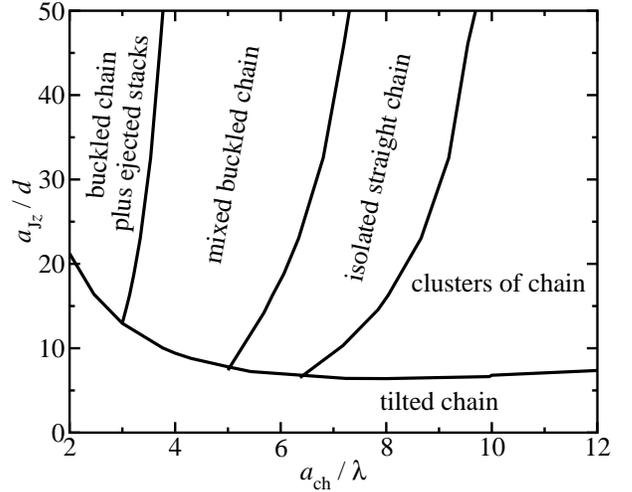}}}
\vspace{-0.2cm}
\caption{ Phase diagram of the isolated vortex chain, showing four phases:
Isolated chain of tilted stacks; Isolated chain of straight stacks crossing
Josephson vortices; Isolated buckled chain; state with chains 
plus ejected vortices. Axes show the spacing of pancake vortices $a_{\rm ch}$
and the Josephson vortex spacing $a_{{\rm J}z}$ within the chain.
  }
\label{fig:three-lions}
\end{figure}

\begin{figure}
\centerline{\resizebox{8cm}{!}{\includegraphics{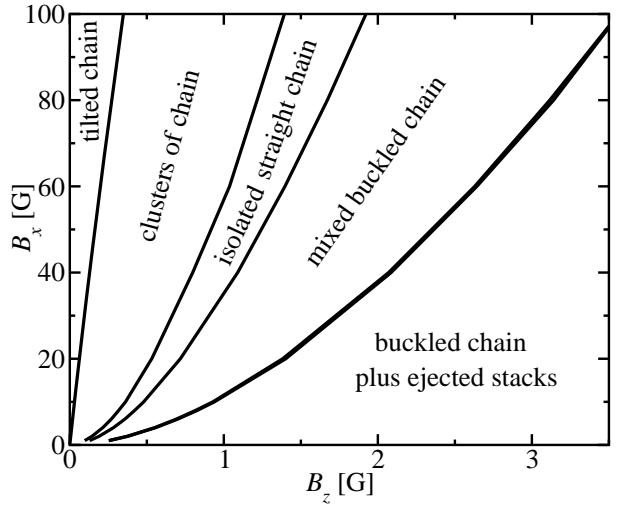}}}
\vspace{-0.2cm}
\caption{ Same phase diagram as in Fig.~2 but with axes
converted to the the in-plane and out-of plane components
of magnetic field, assuming the Josephson vortices form a regular stretched
triangular lattice.
  }
\label{fig:ph-diag}
\end{figure}

In this work we will concentrate on very small out-of-plane fields 
and $ B_z\ll B_x<\Phi_0/\lambda^2$
when all  of the pancake stacks become attached to the centers of
the Josephson vortices, giving the
isolated-chain state observed in 
recent scanning Hall probe measurements.\cite{Grigorenko}
We find that the
stability of the isolated-chain state is quite delicate. 
We first note in Section~\ref{sec:chain}
that the existence of a stable crossing configuration in 
the isolated chains is only possible for large enough anisotropy, layer
spacing, and pancake separation.
We also find that the energy of the isolated chain as a function of 
pancake-stack separation has a minimum, i.e., there is an optimum density
of pancake stacks along a chain, and there will be an instability
towards {\em clustering} when the total density is low.
As the density
of pancake stacks along the chain increases (at a fixed density of
Josephson vortices), the chain will {\em buckle} 
(Section~\ref{sec:buckling}), and then {\em eject} vortices at
a critical minimum separation (Section~\ref{sec:isolated}).
More surprisingly, at very low pancake density the chain state may be replaced
by a chain of tilted stacks
(Section~\ref{sec:tilted}).
These results are summarized in Figs.~\ref{fig:three-lions} 
and~\ref{fig:ph-diag}, where we plot the calculated phase boundaries for
three first-order transitions: clustering, 
buckling
and tilting as well as 
a continuous ejection transition.
Finally, in Section~\ref{sec:disc}
we discuss the extent to which these transitions have been observed
experimentally, and consider the 
effect of fluctuations due to finite
temperature or quenched pinning disorder.

We conclude this introduction with a note on parameters. The experiments we
have referred to,\cite{Bolle,Grigorieva,Grigorenko} were performed 
on the extremely anisotropic cuprate  Bi$_2$Sr$_2$CaCu$_2$O$_x$ (BiSCCO).
Where we make explicit calculations we will take the following appropriate
parameters $\gamma=500$, $\lambda=2000$~{\AA}, and $d=15$~{\AA}. This gives
a Josephson length (the size of the non-linear core of a Josephson vortex)
of $\gamma d=7500$~{\AA}~$\gg\lambda$.

\section{Structure of the Isolated Vortex Chain}
\label{sec:chain}

Koshelev has estimated the structure of a pancake stack that crosses
the center of a single Josephson vortex.\cite{Koshelev99}
In the crossing configuration there is a Lorentz force on the pancakes
in a direction parallel to the Josephson vortex, which must balance
the attraction of each pancake to its stack. 
This distorts the pancake stack, with the largest displacements by the two
pancakes immediately above and below the Josephson core.
The distortions give an energy gain of 
$\Delta E_\times \approx - 8.4 \varepsilon_0 d (\lambda/\gamma d)^2/\ln(3.5
\gamma d/\lambda)$. Here $\varepsilon_0 d=(\Phi_0/4\pi\lambda)^2d$ is the 
typical energy scale for pancake interactions. This
 causes the pancake stack to be attracted to a 
stack of Josephson vortices within the crossing-lattice state.
Koshelev's result uses a quadratic approximation for the 
pancake-to-stack attraction. Recent work\cite{Dodgson_Crete}
using the correct potential
has shown that,
while Koshelev's estimate for the crossing energy is close to the
full result, a stable crossing configuration only exists for extreme
enough anisotropy, $\gamma > \gamma_{\rm min}=2.86\lambda/d$.
In this section we include the interaction between
pancakes in different stacks separated by $a_{\rm ch}$ along an isolated
chain.
For finite $a_{\rm ch}$ we  find a further reduction of
the stability limit of crossing found in 
Ref.~\onlinecite{Dodgson_Crete}. We also find a minimum
in the energy of the chain as a function of stack separation, which 
leads to a clustering instability when $a_{\rm ch}\gg\lambda$.

To calculate the crossing configuration of the chain, 
we use the following assumptions: 
First, we neglect the effect of induced Josephson currents
due to displacing pancakes from their stacks (reasonable for large
values of $\gamma d/\lambda$ and small enough displacements $u<\gamma d$).
Second, we utilize the long range of
the remaining electromagnetic pancake interactions,\cite{Pancakes} 
of order $\lambda$ in the
$z$-direction. There are many ($\sim\lambda/d\sim 10^2$) pancakes
that contribute to the current distribution in one layer, determining
the potential felt by a given pancake. We can therefore reduce this many-body
optimization problem to a one-dimensional problem, considering the
displacement of a row of pancakes in one layer under the potential
due to the ideal chain of pancake stacks (the corrections due to the 
displacements in other layers is small when the number of pancakes
with large displacements is much less than $\lambda/d$).

Within this scheme, the energy profile (per stack) for displacing by $u_n$
a row of pancakes in the $n$th layer when there is a Josephson vortex 
between layers $0$ and $1$ crossing the chain
is,
\be
\Delta E_n(u_n)=
-\frac {\Phi_0 d}{c} J_n^y u_n+V^{\rm row}_{\rm em}(u_n).
\ee
The first term here comes from the Lorentz force on the pancakes due to the 
in-plane
current density from the Josephson vortex, $J_n^y$, and tends to pull the
pancakes away from their stacks. The form of this current
will be discussed more in Section~\ref{sec:buckling}, but the numerical value
is taken from Ref.~\onlinecite{Koshelev_preprint}, e.g.,
the current in the $n=1$ layer (immediately above the Josephson
vortex) is $J_1^y=2.28\varepsilon_0 c/ \Phi_0\gamma d$.
The second term  is the attractive magnetic interaction of the pancake row with
the remainder of all the stacks in the chain.
Clem showed for a single pancake,\cite{Clem}
that this has the same interaction energy as the sum of a pancake
with a  full stack, $V_{\rm em}^{\rm stack}(R)=2\varepsilon_0 d K_0(R/\lambda)$
plus a pancake with its anti-image 
$V_{\rm em}^{\rm pc-pair}=2\varepsilon_0 d \ln(R/L)$, so that for the entire row
we find
\bea
V^{\rm row}_{\rm em}(u)&=&2\varepsilon_0 d \left[
K_0\left(u/\lambda\right) +\ln(u/2\lambda) +\Euler\right]\\
&&\hspace{-2cm}+2\varepsilon_0 d\sum_{j\ne 0}
K_0\left[\frac{ja_{\rm ch}+u}{\lambda}\right]
-K_0\left(\frac{ja_{\rm ch}}{\lambda}\right) +\ln\left[
\frac{ja_{\rm ch}+u}{ja_{\rm ch}}\right].\nonumber
\eea
[Euler's constant $\Euler$
is needed to fix the zero of energy to that of
fully aligned stacks, c.f. the small
$x$ expansion of the modified Bessel function $K_0(x)$].

\begin{figure}
\centerline{\resizebox{8cm}{!}{\includegraphics{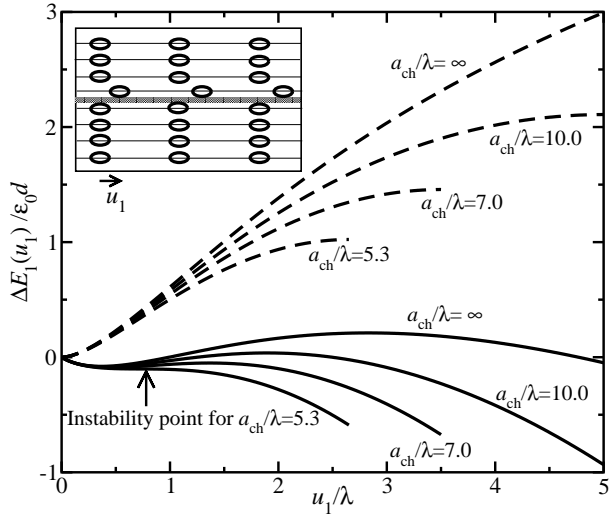}}}
\vspace{-0.2cm}
\caption{ Energy profiles for displacing a pancake row
within a chain of pancake stacks, $V_{\rm em}^{\rm row}(u_1)$, 
(dashed lines) and
the total energy change $\Delta E_1(u_1)$
with a Josephson vortex below the pancake row
 (full lines), for different values of the pancake-stack
spacing $a_{\rm ch}$. Note how the metastable minimum disappears for
spacings smaller than $a_{\rm ch}^{\rm min}=5.3\lambda$. 
Each curve is only plotted
up to $u_1=a_{\rm ch}/2$. The fact that, for all values
of $a_{\rm ch}$ the lowest energy is for large displacements, is not a
physical result, as the Lorentz force argument will break down when $u_1$
is a significant fraction of $a_{\rm ch}$. These results are for
a fixed value of $\gamma d/\lambda=3.75$.
  }
\label{fig:profile}
\end{figure}

Such energy profiles for the $n=1$ pancake row are shown in 
Fig.~\ref{fig:profile}
for different values of 
$a_{\rm ch}$ with the choice of the ratio $\gamma d/\lambda=3.75$ 
(a reasonable choice for BiSCCO). For dilute chains there is a 
stable minimum (as expected from the results of 
Ref.~\onlinecite{Dodgson_Crete}), which determines
the optimal displacement. However, this
stability disappears once the separation is below the critical value
of $a_{\rm ch}^{\rm min}=5.3\lambda$.
 There can be no stable isolated-chain
state for densities higher than this critical value.\cite{relax_caveat} 
Fig.~\ref{fig:profile} also suggests that the crossing
configuration is only metastable, with a lower energy at
$u_1=0.5 a_{\rm ch}$, but this is not reliable as the
simple Lorentz force argument does not hold for pancake displacements of
such a large fraction of the stack-separation. For such displacements
the original Josephson vortex becomes completely fragmented, and a different
approach is needed. In fact, the competing state here is a ``tilted chain'',
and in Section~\ref{sec:tilted} 
we will calculate the energy of the tilted chain,
to compare to the isolated chain state. (A third possibility is a soliton-like
structure\cite{Koshelev_private} that we will not consider here.)

The displacements of the entire stacks are shown in Fig.~\ref{fig:chain} for
$a_{\rm ch}=10\lambda$, and a Josephson-vortex separation of 
$a_{{\rm J}z}=20 d$ (this corresponds to $B_x=53$~G for our parameters).
All of the displacements contribute to the
total crossing energy gain. We find that the
crossing energy per pancake stack only lowers from 
$\Delta E_\times(\infty)\approx -0.21\varepsilon_0 d$ 
for an isolated stack to $\Delta E_\times(a_{\rm ch}^{\rm min})
\approx -0.26\varepsilon_0 d$
as the stacks reach the critical separation $a_{\rm ch}^{\rm min}$.
Therefore within the region of stable crossing, there is a pinning energy
to the Josephson vortex centers per unit length of pancake stack,
\be
\varepsilon_{\rm Jv-pin}=\Delta E_\times/a_{{\rm J}z}\approx
0.2 \varepsilon_0 (d/a_{{\rm J}z}),
\ee
where the last result is with our parameters for BiSCCO.

\begin{figure}
\centerline{\resizebox{8cm}{!}{\includegraphics{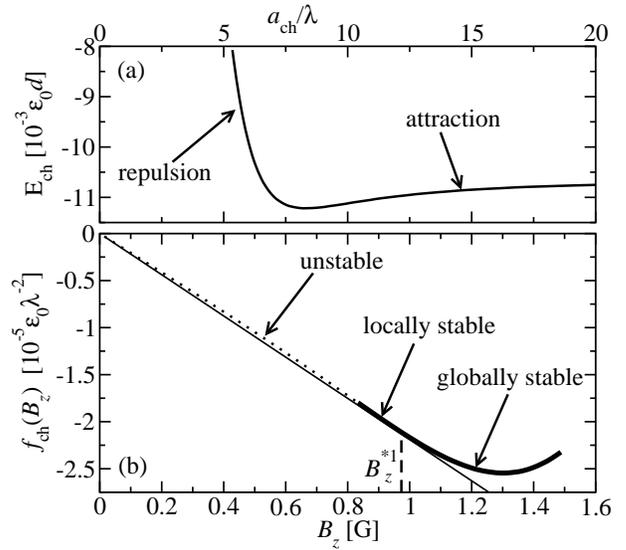}}}
\vspace{-0.2cm}
\caption{ (a) The energy per pancake within the isolated chain  as
a function of separation $a_{\rm ch}$ when $a_{{\rm J}z}=20 d$. 
Note the presence of a minimum energy at
$a_{\rm ch}=8.2\lambda$.
(b) The same result, but expressed as the energy density 
$f_{\rm ch}(B_z)$ 
as a function of the out-of-plane flux density 
$B_z=\Phi_0/a_{\rm ch}a_{{\rm J}y}$. 
The thick line shows where the function has  positive curvature,
required to ensure local stability. The entire function must be convex, and so
for densities less than  $B_z^{*1}$ the chain will phase separate 
into clusters with $a_{\rm ch}=a_{\rm ch}^{*1}$ and regions with 
no stacks.
}
\label{fig:chain_energy}
\end{figure}

In Fig.~\ref{fig:chain_energy}a we plot the total energy per pancake of the
isolated chain $E_{\rm ch}=\frac 1 2 \sum_{j\ne 0} V_{\rm em}^{\rm stack}
(ja_{\rm ch}) + (d/a_{{\rm J}z})\Delta E_\times(a_{\rm ch})$. Note that
there is a minimum at $a_{\rm ch}=8.2\lambda$, reflecting the fact that
at large separations the pancake stacks attract each other. This unusual 
feature has been explained by Buzdin and Baladie\cite{Buzdin_Baladie} 
by considering the distorted
pancake stacks of Fig.~\ref{fig:chain} as the superposition of a straight
stack plus a series of pancake--anti-pancake dipoles. The straight stacks
give a repulsive term, but this is exponentially small 
for $a_{\rm ch}\gg\lambda$,
while the dipoles give a weak attractive term that only falls off like
$1/a_{\rm ch}^2$
(note there is some similarity to the attraction between tilted 
flux lines\cite{TiltedChains}).
Therefore there is a net attraction at large distances, which
will destabilize a homogeneous chain at very low densities.

To formally describe this instability we should consider the free energy
density as a function of out-of-plane field, $f(B_z)=f_{\rm ch}(B_z)+
\varepsilon_{\rm stack}B_z/\Phi_0$. Here $\varepsilon_{\rm stack}$ is the 
line energy of a pancake stack, $f_{\rm ch}(B_z)=E_{\rm ch}(a_{\rm ch})
/a_{\rm ch}a_{{\rm J}y}d$ and $B_z=\Phi_0/a_{\rm ch}a_{{\rm J}y}$.
In Fig.~\ref{fig:chain_energy}b we plot $f_{\rm ch}(B_z)$ using the same
data in Fig.~\ref{fig:chain_energy}a. We also note that the thermodynamically
stable phase is determined by the minimum Gibbs free energy
$g(H_z)=f[B_z(H_z)]-B_zH_z/4\pi$ 
with $H_z=4\pi\partial f/\partial B_z$. By a geometric
construction we see that the point $B_z=B_z^{*1}$
on $f_{\rm ch}(B_z)$ where a straight line
from the origin connects with the same gradient must have $g=0$, and there
is a first-order phase transition between a Meissner phase with no pancake
stacks  and a finite density of stacks with $a_{\rm ch}=a_{\rm ch}^{*1}$. 
All the points
at lower density $B_z<B_z^{*1}$ have $g>0$ and so are thermodynamically 
unstable.
In fact, similar arguments determine that $f_{\rm ch}(B_z)$ must always
be a convex function. The dotted line represents the region where the curvature
of $f_{\rm ch}(B_z)$ is positive, meaning that the solution here is always
unstable. 

The straight line that determines $B_z^{*1}$ is given by 
$f=B\delta H_{\rm c1}^z/4\pi$ where $\delta H_{\rm c1}^z$ is the change in the
lower critical field due to the attraction of pancake stacks to the Josephson
vortices. In Fig.~\ref{fig:chain_energy}b it is given by
$\delta H_{\rm c1}^z=-2.5\cdot 10^{-3} H_{\rm c1}^z$ 
which is hardly measurable.
In real experiments however, the geometry of the samples often makes 
demagnetization effects important such that the average $B_z$ becomes fixed to
the external field $H_{\rm ext}$, 
rather than to $H$. This means that small values of $B_z$
are accessible, and that for $B_z<B_z^{*1}$ we may expect a coexistence of
$B_z=0$ and $B_z=B_z^{*1}$ phases (c.f.\ the intermediate state
in type-I superconductors\cite{LandauLifshitz}).
Alternatively we could describe this mixed regime in terms of ``clusters''
of pancake stacks with separation $a_{\rm ch}^{*1}$. The relative proportion
of space taken up by these clusters is determined  by the value of $B_z$, but
the size of individual clusters
depends on the energy of the ``domain wall'' between the 
cluster of stacks and the region of no stacks, compared to the magnetic energy
cost of large clusters with the wrong flux density.
Also, in the experimental situation one can have different pancake densities 
on the different chains, so the inhomogenous state may have coexisting empty
and filled chains.
Our results for the critical field 
separating the clustered phase from the homogeneous isolated straight chains
are shown in Figs.~\ref{fig:three-lions}
and~\ref{fig:ph-diag}.

\section{Buckling Instability within Vortex Chain}\label{sec:buckling}

It is important to realize that, while the isolated-chain state
gains
energy due to the Lorentz force of the Josephson vortex currents, there
is an energy penalty for the pancake stacks to be close to each other. 
One consequence of this is a maximum density 
above which pancake stacks will be ejected from the chain.
This critical density will be derived in the next section. Below this density,
however, the chain can already react to 
the stack repulsion by buckling.
Note that we will assume that the Josephson vortices remain straight, as in
experiments they are held in place by the interactions with Josephson vortices
in neighboring chains, and the chain separation is much smaller than the
interaction range for Josephson vortices, $a_{{\rm J}y}\ll\gamma\lambda$.
In contrast, the interactions between pancake stacks on different chains
are in the dilute limit, $a_{{\rm J}y}\gg \lambda$, and so are easily 
displaced.

To calculate this buckling, we need to know the energy gain 
from a crossing event
when the pancake stack is displaced away from the center of a
Josephson vortex. We therefore need the full current profile of a
Josephson vortex, which is a 
solution of the non-linear equations that arise from the 
London-Lawrence-Doniach model.\cite{Jvortex,Bulaevskii92}
An accurate numerical solution for a Josephson vortex is described in
Appendix~B of Ref.~\onlinecite{Koshelev93}. For a vortex directed along
$\hat{\bf x}$ centered at $y=0$ and between the $n=0,1$ layers, the
current in the $y$ direction can be written\cite{Koshelev_preprint}
(ignoring screening, i.e. $n<\lambda/d$ and $y<\lambda_c$),
\be
J_n(y)=\frac{2 c \varepsilon_0}{\Phi_0 \gamma d} p_n(y/\gamma d),
\ee
with $p_n(\tilde{y})=\phi_n'(\tilde{y})$ the reduced superfluid momentum,
where the phase has the form,\cite{Koshelev2000}
\bea
\phi_n(\tilde{y})& =&
\tan^{-1}\left[\frac{(n-\frac 12)}{\tilde{y}}\right]
+ 
\frac{0.35 (n-\frac 12)\tilde{y}}{[(n-\frac 12)^2 + \tilde{y}^2 + 0.38]^2}
\nonumber\\
&&+
\frac{8.81 (n-\frac 12)\tilde{y} [\tilde{y}^2 - (n-\frac 12)^2 + 2.77]}
{[(n-\frac 12)^2 + \tilde{y}^2 + 2.02]^4}.
\eea
When $y=0$, this gives the results used in Section~\ref{sec:chain},
with $J_n(0)=(2 c \varepsilon_0/\Phi_0 \gamma d) C_n/|n-\frac 12|$, and
$C_1=0.57$, 
$C_2=0.86$, 
$C_3=0.99$, and $C_n\approx 1$ for $n>3$.

We can now recalculate the crossing energy for an arbitrary distance between
a pancake stack and the center of a Josephson vortex. Within the quadratic
approximation used in 
Refs.\onlinecite{Koshelev99}~and~\onlinecite{Koshelev_preprint}
one can easily solve the crossing configuration for a single stack. 
Writing $\Delta E_n(u_n)= - (\Phi_0d/c)J_n(y) u_n +\frac 12 \alpha u_n^2$,
where we take $\alpha=(\varepsilon_0 d/\lambda^2)\ln(\lambda/r_w)$ 
with $r_w$ a short distance cut-off, we find the relaxation energy,
\be\label{eq:Exofy}
\Delta E_\times(y)=-2\left(\frac {\lambda}{\gamma d}\right)^2
\frac 1 {\ln(\lambda/r_w)} \sum_{n=-\infty}^\infty 
\left[p_n(y/\gamma d)\right]^2.
\ee
Koshelev takes $r_w=u_1$ ($\approx 0.29\lambda^2/\gamma d$ for $y=0$), 
and in 
Ref.~\onlinecite{Dodgson_Crete} we have shown this to be a good approximation
to the result when the full form of the pancake-stack interaction is used.

\begin{figure}
\centerline{\resizebox{8cm}{!}{\includegraphics{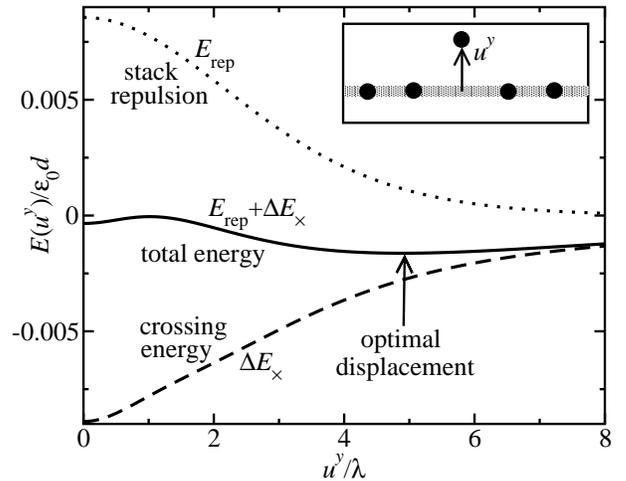}}}
\vspace{-0.2cm}
\caption{ The energy profile when we pull one pancake stack away 
a distance $u^y$ from the isolated chain state, for $a_{\rm ch}=5.5\lambda$, 
and $a_{{\rm J}z}=25 d$. The geometry is shown in the inset. The dashed line
shows the energy from the crossing events on the displaced stack. The
dotted line is the interaction energy of displaced stack with the other
stacks on the chain. The full line is the sum of these two contributions.
Note the presence of two minima, one for $u^y=0$, and a lower minimum at 
$u^y=4.9\lambda$.
  }
\label{fig:one_displaced}
\end{figure}
This result for the crossing energy as a function of stack displacement $u^y$ 
is
shown as the dashed line in  Fig.~\ref{fig:one_displaced}, where there is
one Josephson vortex for every 25 pancakes along a stack.
As might be expected, the energy increases quadratically
at small $u^y$. More interesting is the fact that the crossing energy vanishes
only as $\Delta E_\times\approx - \lambda^2/{u^y}^2$ for $u^y\gg\gamma d$ 
(note that
an exponential suppression will occur at extremely long distances
$u^y>\lambda_c$). In contrast, also shown in Fig.~ \ref{fig:one_displaced}
(the dotted line) for $a_{\rm ch}=5.5\lambda$
is the repulsive energy cost
for displacing one pancake stack away from a chain,
$
E_{\rm rep}=\sum_{j\ne 0} V_{\rm em}^{\rm stack}\left(\sqrt{(ja_{\rm ch})^2
+{u^y}^2}/\lambda\right).
$ 
This contribution vanishes exponentially when 
$u^y>\sqrt{a_{\rm ch}\lambda}$, and so the total energy is always negative
for large enough $u^y$. This gives the possibility of two minima
in the total energy (see the full line in Fig.~\ref{fig:one_displaced}), 
and a first-order transition where the displacement jumps
to a finite value as $a_{\rm ch}$ decreases.
While we have started by considering
displacing a single stack from the chain,
it is this instability that drives a buckling of the whole chain.

\begin{figure}
\centerline{\resizebox{8cm}{!}{\includegraphics{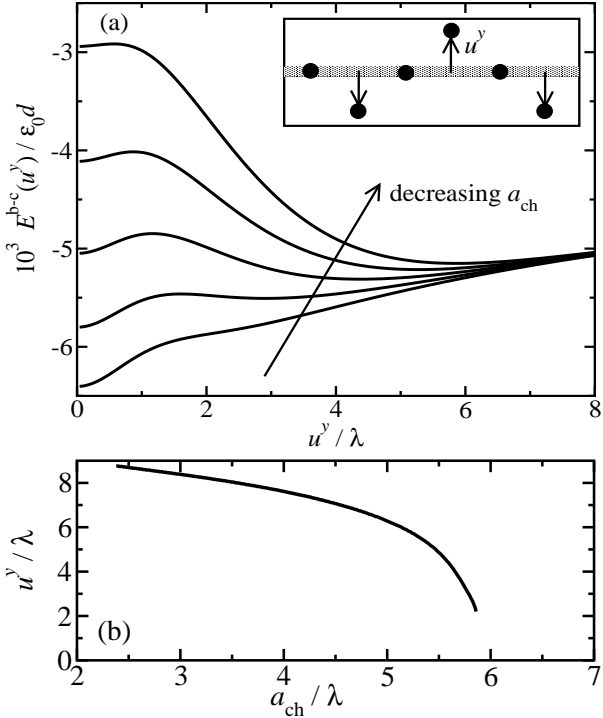}}}
\vspace{-0.2cm}
\caption{(a) Energy of the buckled chain (see inset) as a function of the
buckle parameter $u^y$, for stack separation along the chain
$a_{\rm ch}/\lambda=5.2$, $5.4$, $5.6$, $5.8$, and $6.0$. For the separation
$a_{\rm ch}/\lambda=6.0$ there is only one minimum at $u^y=0$.
For $a_{\rm ch}/\lambda=5.8$ there is a second minimum at $u_y=3.3\lambda$.
The remaining cases have their lowest minimum at finite $u^y$.
(b)~The size of the optimum buckling distortion $u^y_{\rm opt}$ as a function
of $a_{\rm ch}$.
  }
\label{fig:buckle_two}
\end{figure}
To estimate the buckling transition we use a variational approach for the
buckled configuration (See the inset of Fig.~\ref{fig:buckle_two}a). 
This configuration is only characterized
by a single displacement parameter $u^y$, and so we can look at the energy 
profiles in $u^y$ for different $a_{\rm ch}/\lambda$. First, the energy
per pancake is calculated before crossing relaxations,
\be
E^{\rm b-c}_0(u^y)= \frac{\varepsilon_0 d}{N} \sum_{i\ne j}
 K_0\left(\sqrt{(i-j)^2 a_{\rm ch}^2 + (y_i-y_j)^2}
/\lambda\right)
\ee
with $y_i=0$ for $i$ even, and $y_i=(-1)^{(i+1)/2}u^y$ for $i$ odd.
Next we checked for the stability of crossing, as in Section~\ref{sec:chain},
and found that the crossing configuration becomes unstable if 
$a_{\rm ch}<2.65 \lambda$. Finally, to calculate the energy gain from crossing
the Josephson vortices, we use the result (\ref{eq:Exofy}) giving the total
energy of the buckled chain,
\be
E^{\rm b-c}(u^y) = E^{\rm b-c}_0(u^y) +
\frac{d}{2a_{{\rm J}z}}[\Delta E_\times(0)+\Delta E_\times(u^y)].
\ee
In Fig.~\ref{fig:buckle_two}a
we plot this energy for $a_{\rm ch}/\lambda$ from $5.2$ to $6.0$,
with $a_{{\rm J}z}=25d$,
showing how the minimum energy of a buckled chain crosses the energy of a 
straight chain as $a_{\rm ch}$ decreases. 
Note that the minimum for a finite $u^y$ is quite shallow, 
and so we might expect large fluctuations in the extent of buckling 
along the chain due to random vortex pinning, which is discussed in 
Section~\ref{sec:disc}.
Also shown in Fig.~\ref{fig:buckle_two}b
 is the optimal displacement $u^y_{\rm opt}$ as a function of
$a_{\rm ch}$, which has a smallest value of   $u^y_{\rm opt}=2.3\lambda$
at the first appearance of (meta)stable buckling at $a_{\rm ch}= 5.8\lambda$ 
and then increases at higher densities.

\begin{figure}
\centerline{\resizebox{8cm}{!}{\includegraphics{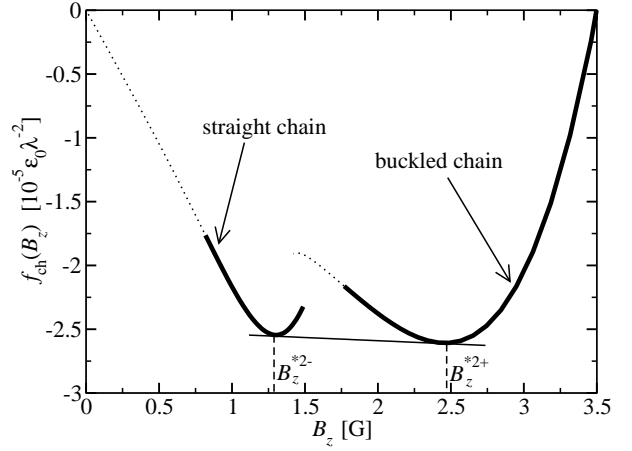}}}
\vspace{-0.2cm}
\caption{
Energy density of the buckled-chain state with $a_{{\rm J}z}=20 d$
as a function of out-of-plane
field $B_z$. Also shown is the result for the straight chain from Fig.~5b.
The two points joined by the straight line have the same Gibbs free energy,
and those densities between these points are thermodynamically unstable towards
a mixed phase of straight and buckled chains.
  }
\label{fig:buckle_transition}
\end{figure}

It has been shown in Section~\ref{sec:chain} that one should plot the energy
density of the chain as a function of the out-of-plane flux density $B_z$,
in order to determine
the stable thermodynamic phases. Such a plot is shown in 
Fig.~\ref{fig:buckle_transition} comparing the
straight and buckled chains. It shows that the isolated straight chains at
$B_z=B_z^{*2-}\approx 1.3$~G has the same Gibbs free energy as the 
buckled-chain state at $B_z=B_z^{*2+}\approx 2.5$~G. Therefore at fields with
$B_z^{*2-}<B_z<B_z^{*2+}$ the thermodynamic phase will be a coexistence of
straight and buckled regions, with the relative proportion linearly dependent
on $B_z$. Note that $B_z^{*2-}$ is only slightly higher than 
$B_z^{*1}$ so that there is only a small regime where pure isolated
straight chains are the stable phase.
The same procedure has been followed for a range of $a_{{\rm J}z}/d$, and the
resulting boundaries to the mixed buckled chains are
shown in Figs.~\ref{fig:three-lions} and~\ref{fig:ph-diag}.

\section{Ejection of Pancake Stacks from Isolated Chain}\label{sec:isolated}

We now consider a simple model for the energy of the composite state where 
a fraction of the pancake stacks are not located on the chains. This model
will show that there is a continuous phase transition between this composite
state and the isolated-chain state as a function of $B_z$.
The order parameter of this transition is the density of the dilute lattice
of pancake stacks not trapped on chains. The simple model assumes that the
chain spacing $a_{{\rm J}y}$, and the spacing between pancake stacks is much 
greater than the penetration depth $\lambda$. In this case we can separate the
total interaction energy density of pancake stacks in the form 
$f_{\rm tot}=f_{\rm ch}(\nu_{\rm ch})+
f_{\rm dil}(\nu_{\rm dil})$, where $\nu_{\rm ch}$ and
$\nu_{\rm dil}$ are the (2D) densities of pancake stacks on and off the chains
respectively, and the total density is $\nu_{\rm ch}+\nu_{\rm dil}=
\nu_{\rm tot}=B_z/\Phi_0$.
The separation of pancakes along the chain is 
$a_{\rm ch}=1/a_{{\rm J}y}\nu_{\rm ch}$.
The chain energy $f_{\rm ch}(\nu_{\rm ch})$ is to be calculated as in
Sections~\ref{sec:chain} and~\ref{sec:buckling} for straight and buckled
chains respectively.
We will see that the phase transition to the composite state occurs
when there is a minimum in $f_{\rm ch}(\nu_{\rm ch})$, and so near this
transition we will expand,
\be
f_{\rm ch}(\nu_{\rm ch})=
\epsilon'\lambda^2 (\nu_{\rm ch}-\nu_{\rm ch}^{\rm c})^2.
\ee

The density of pancake stacks not on a chain is small near the transition,
so that the interaction energy density is simply (using the limit
$K_0(x)\approx\sqrt{\pi/2x}e^{-x}$ and only including nearest neighbors),
\be
f_{\rm dil}(\nu_{\rm dil})
=3 \varepsilon_0\sqrt{2\pi\lambda}\nu_{\rm dil}^{-5/4}
e^{-1/\lambda\nu_{\rm dil}^{1/2}}.
\ee
Note that in the small $\nu_{\rm dil}$ limit, all terms in a power series
expansion of $f_{\rm dil}(\nu_{\rm dil})$ are zero, i.e. the function is
extremely flat. For this reason the critical total density at the transition
is only determined by the minimum in $f_{\rm ch}(\nu_{\rm ch})$
at $\nu_{\rm ch}^{\rm c}$. For $\nu_{\rm tot}<\nu_{\rm ch}^{\rm c}$, 
all of the pancake
stacks are on chains. At densities just above $\nu_{\rm ch}^{\rm c}$
the energy is minimized with a dilute off-chain density,
\be
\nu_{\rm dil}=\delta \nu_{\rm tot} - \frac{3\sqrt{2\pi}}{4}
\frac{\varepsilon_0}{\varepsilon'}\frac 1 {\lambda^{5/2}
\delta \nu_{\rm tot}^{1/4}}
e^{-1/\lambda\delta \nu_{\rm tot}^{1/2}}.
\ee
The second term is extremely small for the fields we are interested in
 $B_z\ll\Phi_0/\lambda^2$, so that we can say in the composite state at
$\nu_{\rm tot}>\nu_{\rm ch}^{\rm c}$ there is a
fixed density on the chains $\nu_{\rm ch}=\nu_{\rm ch}^{\rm c}$ and the remaining
density is $\nu_{\rm dil}=\nu_{\rm tot}-\nu_{\rm ch}^{\rm c}$. 
The phase boundary $\nu_{\rm ch}^{\rm c}$ between the composite state and the 
isolated-chain state is calculated from the minimum of $f_{\rm ch}$.
In Figs.~\ref{fig:three-lions}
and~\ref{fig:ph-diag} this boundary is plotted, but is not distinguishable by
eye from the transition between mixed buckled and pure buckled chain. Therefore
the pure buckled chain has a very narrow range of existence.
It is worth stating that an ejection transition has been observed 
experimentally\cite{Grigorenko} with
results that seem to be consistent with the above, 
although a quantitative analysis of the transition has
not yet been published.

\section{Crossing-to-Tilted transition within isolated chains}
\label{sec:tilted}

\begin{figure}
\centerline{\resizebox{8cm}{!}{\includegraphics{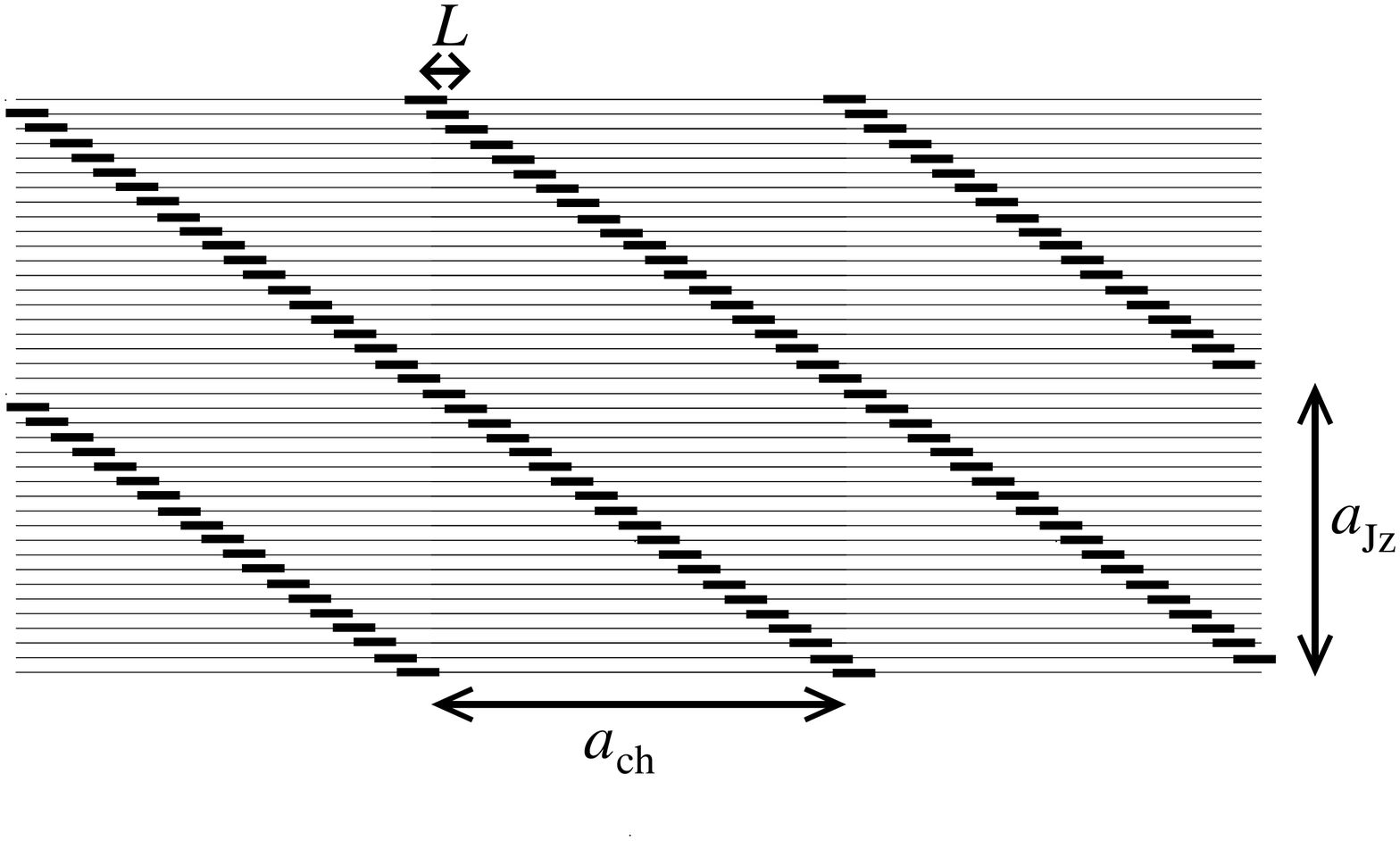}}}
\vspace{-0.2cm}
\caption{
Configuration of the tilted chain, where pancakes in a stack are displaced from
layer to layer by $L=a_{\rm ch}/a_{{\rm J}z}$.
  }
\label{fig:tilted_chain}
\end{figure}

In this section we calculate the energy of a chain of tilted vortices, and 
compare to the chain of crossing vortices.
The tilted chain is
an alternative configuration to the crossed chain
with the same density of pancake and 
Josephson vortices
(see Fig.~\ref{fig:tilted_chain}) where the tilt
angle is determined by $\tan\theta=L/d = a_{\rm ch}/a_{{\rm J}z}$.
These tilted stacks have an increased energy of pancake interactions 
but there may be an energy gain from replacing fully developed Josephson
vortices with many short Josephson segments of length $L<\gamma d$.
To calculate the energy of the tilted chain, we use the linear approximation
of the London-Lawrence-Doniach model,\cite{Bulaevskii92,Koshelev93} 
where we ignore the non-linear
effects from regions with large phase differences across neighboring layers.
This approximation is justified as long as the length $L$ is much smaller than
the Josephson length $\gamma d$ and the pancake separation $a_{\rm ch}$.
Note that the condition $L<\gamma d$ corresponds to $B_x/B_z<\gamma$, which is
the case we are interested in.
The energy due to pancake interactions in the linear approximation is,
\be
E_{\rm pc}^{\rm tilt}
=\frac{1}{8\pi}\int \frac{d^2k}{(2\pi)^2}\frac{dq}{2\pi}
\frac { \left[ 1+(\gamma\lambda)^2(k^2 +\tilde{q}^2)\right]\,\, |S_z|^2}
{  \left[ 1+\lambda^2(k^2 +\tilde{q}^2)\right]
\left[ 1+(\gamma\lambda)^2k^2 + \lambda^2\tilde{q}^2\right]
}
\ee
with $S_z({\bf k},q)=\Phi_0 d \sum_{j,n} e^{-iqnd}e^{-ik_x(ja_{\rm ch}+nL)}$
and  $\tilde{q}^2=(2/d)^2\sin^2(qd/2)$.
The integral should be cut off due to the pancake cores at $k>\pi/\xi_{\rm ab}$
and due to the layered structure at $q>\pi/d$.
Similarly, the Josephson segments interact with an energy contribution,
\be
E_{\rm Jv}^{\rm tilt}
=\frac{1}{8\pi}\int \frac{d^2k}{(2\pi)^2}\frac{dq}{2\pi}
\frac { |S_x|^2}
{
\left[ 1+(\gamma\lambda)^2k^2 + \lambda^2\tilde{q}^2\right]
}
\ee
with
$S_x=\frac{2\Phi_0}{k_x} \sin\left(\frac{k_xL}{2}\right) 
\sum_{j,n} e^{-iq(n+\frac 12)d}e^{-ik_x[ja_{\rm ch}+(n+\frac 12)L]}$.

\begin{figure}
\centerline{\resizebox{8cm}{!}{\includegraphics{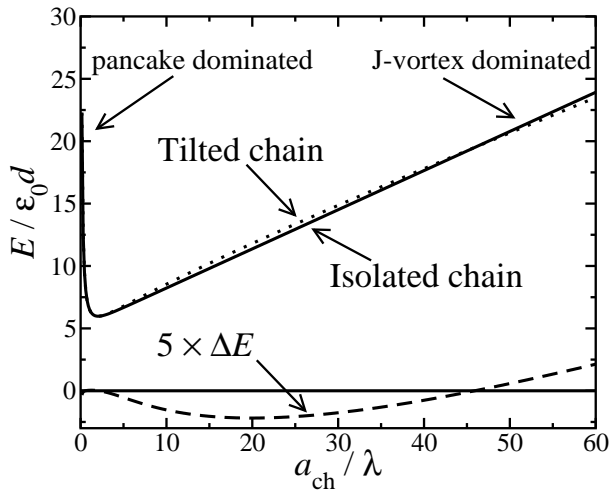}}}
\vspace{-0.2cm}
\caption{
Energy per pancake of the tilted chain (dotted line) and the chain of straight
pancake stacks crossing Josephson vortices, for $a_{{\rm J}z}=25 d$, plotted
against the pancake separation $a_{\rm ch}/\lambda$. Also shown (dashed line)
is the difference of the two energies, which crosses zero at
$a_{\rm ch}=45\lambda$.
  }
\label{fig:tilt_vs_chain}
\end{figure}

The integrals can all be done exactly, leaving a sum over the values of
$k_x=2\pi m/a_{\rm ch}$. We have evaluated these sums numerically to find the
energy of the tilted chain. This is shown for $a_{{\rm J}z}=25d$
as the dotted line in 
Fig.~\ref{fig:tilt_vs_chain}.
Also shown as the full line
is the energy per pancake of the crossing chain with the same
density of pancakes and Josephson vortices, $E_{\rm cross}=
E_{\rm ch} + E^{\rm cross}_{\rm Jv}$, where 
$E_{\rm ch}$ is defined in Section~\ref{sec:chain} and
\be
E^{\rm cross}_{\rm Jv}=
\frac{\varepsilon_0 d}{\gamma}
\frac{a_{\rm ch}}{a_{{\rm J}z}}\left[ 1.55 
+\ln\left(\frac\lambda d\right) +\sum_{j\ne 0}K_0 \left(
\frac{ja_{{\rm J}z}}{\lambda}\right)\right].
\ee

The tilted chain has a lower energy from Josephson vortex
contributions (dominant at low density) but a higher energy from
pancake vortex contributions (dominant at high density), and
there is a 
first-order transition between the isolated chain of straight stacks
and the chain of tilted stacks.
In fig.~\ref{fig:tilt_vs_chain} it is shown that for 
$a_{\rm ch}<45\lambda$ (higher density) the preferred state is the crossing 
chain, while for $a_{\rm ch}>45\lambda$ 
(lower density), the lowest energy is for the tilted chain. 
Note that Savelev et al.,\cite{Savelev} also find a ``reentrant'' transition
back to a tilted lattice for fields close to the $ab$-plane, when 
$\lambda>\gamma d$.

A strict 
treatment should follow Sections~\ref{sec:chain}
 and~\ref{sec:buckling} and consider the points of the same
Gibbs free energy density.
This is not possible within our level of treatment, however, as we would need
accurate results at very low $B_z<B_x/\gamma$ where non-linear effects in the
tilted chain are important. Here, we can only show that there must be
a low-density transition to the tilted chain, but we cannot plot the
coexistence boundaries. Even so,
we plot the line where the energies cross in Figs.~\ref{fig:three-lions}
and~\ref{fig:ph-diag} separating the clustered chain from a tilted chain phase.
In real experiments there is a finite 2D density 
of Josephson vortices,
which cannot be considered isolated in the sense of the pancake vortex chains
(the interaction range $\lambda_c$ is greater than the chain spacing
$a_{{\rm J}y}$).
Therefore a full calculation of the 3D tilted lattice
requires a minimization over the
aspect ratio, which  we leave for a future work.

\section{Discussion}\label{sec:disc}

We now discuss the extent to which these transitions have been observed
experimentally. 
In the recent scanning
Hall probe experiments of Grigorenko et al.,\cite{Grigorenko}
there does seem to be a sharp transition as a function of $B_z$
between a composite state
of chains with a dilute lattice and the isolated chain state, as derived
in Section~\ref{sec:isolated}. 
At lower fields there is some evidence for buckling and 
clustering (Sections~\ref{sec:chain} and ~\ref{sec:buckling}) 
although the influence of pinning disorder
may be contributing to this. Finally, Grigorenko et al.\ report a 
strange transition at very low $B_z$  where the chains are replaced by faint,
homogeneous ``stripes'' of flux. These stripes would be consistent with the 
flux distribution from a tilted chain, and it is likely that a transition 
from the isolated-chain state to a tilted-chain state (as found in 
Section~\ref{sec:tilted}) takes place.
Experimentally, a large jump in $B_z$ (i.e. in the pancake density) is seen
at the transition to tilted chains; unfortunately we were not able to calculate
the jump in this paper, as it requires an
accurate treatment of non-linear effects
in the small $B_z$ limit.
Other experimental features are that the tilted chains are much straighter than
the crossing chains, and less pinned. This is consistent with the fact that
the tilted chains are really part of a 3D line lattice with stronger 
interactions that dominate over pinning disorder, whereas in the crossing case 
the pancake stacks are in the extreme dilute limit, and therefore easily pinned
in random low-energy sites.

Finally, we briefly discuss the effect of fluctuations.
The calculations in this paper have found the lowest-energy vortex
configurations of the chains. In reality there will be distortions of these
states due to thermal fluctuations and random pinning to inhomogeneities 
in the underlying crystal. It is well established that thermal fluctuations 
can lead to a melting of the vortex crystal.\cite{Blatterreview} A melting
transition to completely decoupled layers is studied for $B_x=0$ in 
Ref.~\onlinecite{Dodgson_Substrate}, involving short-wavelength fluctuations
of pancake stacks with $k_z>\pi/\lambda$.
While there will be some modification in the presence of Josephson vortices
in tilted fields, in the low out-of-plane fields studied here, 
$B_z\ll\Phi_0/\lambda^2$, this melting will take place at a high temperature,
close to the Berezinskii-Kosterlitz-Thouless transition\cite{BKT} at
$T_{\rm BKT}=\varepsilon_0 d/2$.
If we consider long wavelength fluctuations, $k_z\rightarrow 0$, then we may
expect some kind of melting of the chain at very low fields due to the 
exponentially small interaction of pancake stacks. However, the mechanism 
of such a transition must be different from the ``entanglement'' proposed
in the original derivation of low-field vortex lattice melting by 
Nelson.\cite{Nelson}
Considering the nature of the crystalline order of the isolated vortex chain,
we should recognize the two-dimensional nature of this state. A simple 
consequence is that thermal fluctuations at
long wavelength will lead to a quasi-long-range ordered state. It may then
be possible to have a 2D continuous melting transition via the unbinding of
dislocations. However, a dislocation for this system of stacks corresponds to
a stack that terminates in the middle of the chain, and this should cost an
energy linear in the system size, rather than the usual logarithmic energy
of a dislocation. This will suppress a dislocation unbinding transition.

In the experiments of Grigorenko et al.\ the pinning disorder seems to be
more important in disturbing the chain states than thermal fluctuations. The
fact that the pancake stacks are observed in fixed positions is due to pinning
(otherwise thermal fluctuations would smear out the average density in the 
chain state),
and this also tends to disorder the chains. On general grounds we expect 
pinning-induced wandering of the pancake stacks within the chains
so that there is only a short range order. There will also be significant
transverse displacements, and the buckling effect should be enhanced by 
disorder.

It is a pleasure to  thank Simon Bending and Sasha Grigorenko for 
enthusiastically introducing me to the world of chains and crossing lattices,
as well as  Dima Geshkenbein for useful 
discussions and Alex Koshelev for demonstrating how to calculate the
currents near a Josephson vortex.
The author is supported by an EPSRC Advanced Fellowship AF/99/0725.
Work at the ETH-Z\"urich was financially supported by the Swiss National
Foundation.


\end{document}